\documentclass[preprint,aps]{revtex4}

\usepackage{epsfig}
\usepackage{color}

\def\e{\mathop{\rm \mbox{{\Large e}}}\nolimits}

\newcommand{\be}{\begin{equation}}
\newcommand{\ee}{\end{equation}}
\newcommand{\bc}{\begin{center}}
\newcommand{\ec}{\end{center}}
\newcommand{\bea}{\begin{eqnarray}}
\newcommand{\eea}{\end{eqnarray}}
\newcommand{\ba}{\begin{array}}
\newcommand{\ea}{\end{array}}

\begin{document}

\title{Effects of Mass Media action on the Axelrod Model with Social Influence}
\author{Arezky H. Rodr\'{\i}guez$^{1}$, Y. Moreno$^{2,3}$}

\affiliation{$^{1}$ Academia de Matem\'aticas, Universidad Aut\'onoma de la Ciudad de M\'exico,
Mexico City 03020, Mexico. \\
$^{2}$ Instituto de Biocomputaci\'on y F\'{\i}sica de Sistemas Complejos, Universidad de Zaragoza,
Zaragoza 50009, Spain. \\
$^{3}$ Departamento de F\'{\i}sica Te\'orica, Facultad de Ciencias, Universidad de Zaragoza,
Zaragoza 50009, Spain.}

\date{\today}

\begin{abstract}
The use of {\it dyadic interaction} between agents, in combination with {\it homophily} (the principle that
``likes attract'') in the Axelrod model for the study of cultural dissemination has two important problems: the
prediction of monoculture in large societies and an extremely narrow window of noise levels in which diversity with
local convergence is obtained. Recently, the inclusion of {\it social influence} has proven to overcome them (A. Flache
and M. W. Macey, arXiv:0808.2710). Here we extend the Axelrod model with social influence interaction for the study of
Mass Media effects through the inclusion of a super-agent which acts over the whole system and has nonnull overlap with
each agent of the society. The dependence with different parameters as the initial social diversity, size effects, Mass
Media strength and noise is outlined. Our results might be relevant in several socio-economic contexts and for the
study of the emergence of collective behavior in complex social systems.
\end{abstract}



\maketitle

\section{Introduction.}

Social human systems are constituted by human agents which interact with a limited number of peers. This perhaps
apparently random procedure yields to global regularities: there are order-disorder transitions (spontaneous formation
of cultures, the emergence of consensus about different issues, spontaneous formation of a common language), examples
of scaling and universality, to mention just a few. These phenomena are now been under intense studies through
different techniques. One of them is the so called agent-based modelling
\cite{simul-soc-scient,art-soc,grw-art-soc,grimm,robert2005,bonabeau}.

The interest in the application of such techniques has grown rapidly mainly as a result of the increasing availability
of computational capabilities. Today, computer simulations have become an excellent way to model and understand social
processes. The usefulness of social simulation modeling results as much from the process (problem specification, model
development, and model evaluation) as the product (the final model and simulations of social system dynamics).
Nevertheless, social simulations need to be a theory-guided enterprise. Its results will often be the development of
explanations, rather than the prediction of specific outcomes \cite{simul-soc-scient}.

Agent-based simulations allows us the possibility to develop a new way of thinking, parallel to analytical procedures,
based on the idea about the emergence of complex behaviors from relatively simple activities \cite{simul-soc-scient}.
It has great potential to assist in discovery of simple social effects when social scientists build very simple models
that focus on some small aspect of the social world and discover the consequences of their theories in the ``artificial
society'' they have built.

One of those simple models already built is the well known Axelrod model for the study of dissemination of culture
\cite{axelrod-original-paper,axelrod-libro,vespignani}. He argued that culture ``is something people learn from each
other'', and hence something that evolves through social influence. To study the process of cultural diffusion Axelrod
built a model based on two simple assumptions:
\begin{enumerate}
\item people are more likely to interact with others who share many of their cultural attributes, and
\item these interactions tend to increase the number of cultural attributes they share (thus making them more likely to
interact again).
\end{enumerate}

The Axelrod model has been exhaustively implemented to study a great variety of problems: the nonequilibrium phase
transition between monocultural and multicultural states \cite{klemm2003-2}, the cultural drift driven by noise
\cite{klemm2003-1,sanctis}, nominal and metric features \cite{flache,jacobmeier}, propaganda \cite{carletti2006}, time
evolution dynamics \cite{vazquez2007}, the resistance of a society to the spread of foreign cultural traits
\cite{boccara}, finite size effects \cite{toral2006}, the impact of the evolution of the network structure with
cultural interaction \cite{centola2007}, and mobility of social agents \cite{us}, among others.

The dynamical influence on the society is described in the original Axelrod model using a {\it dyad interaction} which, at
each time step, only comprises the source and the target of influence in a particular interaction. As discussed by A.
Flache and M. W. Macy in Ref. \cite{AFlache08082710}, the influence a person feels from the society is a {\it social}
phenomenon that cannot be reduced to the interactions within a dyad given by source-target couple of persons because
the social pressure on the target to adopt an opinion is proportional to the number of people that the target perceives
are supporting this opinion. On the contrary, the Axelrod model assumes that influence is interpersonal (dyadic). We are
then in front of a two completely different assumption with respect to the way people interact in society:
\begin{enumerate}
\item dyadic interaction where it is supposed that two people in a relationship interact in isolation from others and
\item social influence which is multilateral and involves all network neighbors simultaneously.
\end{enumerate}

In Ref. \cite{AFlache08082710} the authors explained that the central implications of Axelrod's model profoundly change
if the {\it dyadic interaction} is changed to {\it social influence}. The authors have shown that the combination of
social influence with {\it homophily} (the principle that ``likes attract'') solves two important problems which
contradict the empirical pattern:
\begin{enumerate}
\item the original Axelrod model predicts cultural diversity in very small societies, but monoculture in larger ones \cite{fontanari-1} and
\item when cultural perturbation is present, diversity is obtained only in a very narrow window of noise level and
this window decreases with increasing population size \cite{klemm2003-1,klemm2003-2}.
\end{enumerate}

On the other hand, one of the extensions of the Axelrod model that have been done consists of the study of Mass Media
effects over the society \cite{shibanai2001,avella2005,avella2006,avella2007}. Surprisingly, the Mass Media has been
found to be a destabilizing factor which drives the system to a multicultural state \cite{fontanari-2}. The exception
are the works \cite{mazzitello2007,mazzitello2008,arezky-ijmpc2009} where different models has been introduced.

The aim of this work is to investigate the effects of Mass Media over the society when the social influence (in the
sense of Ref. \cite{AFlache08082710}) is considered. This paper is organized as follows: in section \ref{axelrod-model}
a brief description of the Axelrod model and its results are explained, in section \ref{the-model} it is introduced the
model which considers social influence, in combination with homophily and Mass Media Our results are analyzed in
section \ref{results} and finally some conclusions are outlined.

\section{\label{axelrod-model} A brief description of the Axelrod model.}

The Axelrod model consists of a population of agents, each one occupying a single node of a square network of size $L$
and area $L \times L$. The culture of an agent is described by a vector of $F$ integer variables $\{ \sigma_f \}$
called {\it features} ($f$ = 1, ..., $F$). Each feature can assume $q$ different values between 0 and $q - 1$. These
are the possible {\it traits} allowed per feature. In the original Axelrod model the interaction topology is regular
bounded (non-toroidal). Each agent can interact only with its four neighbors (a von Neumann neighborhood) which are the
most closer (only one step distant from the target agent of influence) without crossing the borders. Initially,
individuals are assigned a random culture and the parameter $q$, which defines the possible traits in each cultural
dimension, can be seen as a measure of the initial disorder or cultural variety in the system. In the temporal dynamics
of the model at each time step $t$, the cultural profile of a random selected agent $i$ may be updated through the
interaction with a randomly chosen neighbor $j$. The probability of this interaction is proportional to the
corresponding overlap of their cultural profiles (the amount of features with identical traits). When interacting,
agent $j$ influences agent $i$ causing the last to adopt $j$'s trait on a feature randomly chosen from those that they
do not share. The process outlined above continues until no cultural change can occur. This happens when every pair of
neighboring agents have cultures that are either identical or completely different.

In the model it is explored the aggregate behavior by studying the spatial distribution of the emergent cultural
regions: sets of spatially contiguous agents who share an identical vector of culture. Two final possible states are
possible: only one cultural region is obtained or multiple cultures are obtained separated by a boundary. These states
are called monocultural and multicultural (or global, polarized), respectively. A first order phase transition is
obtained for $F >$ 2 between the monocultural to a multicultural state for increasing value of the initial diversity
$q$ \cite{vespignani}.

The Axelrod model has shown how diversity with local convergence can be obtained. Nevertheless, as mentioned before,
there are two results of the model which contradict empirical patterns: (1) the Axelrod model predicts cultural
diversity in very small societies, but monoculture in larger ones and (2) when cultural perturbation is present,
diversity is obtained only in a very narrow window of noise level and this window decreases with increasing population
size \cite{klemm2003-1,klemm2003-2}. This has been justified as a consequence of the dyadic interaction considered
between the agents of the society \cite{AFlache08082710}.

\section{\label{the-model} The model for self-included social influence with Mass Media.}

Following the proposition made by A. Flache {\it et al} in Ref. \cite{AFlache08082710} we have developed a new
interaction procedure in the Axelrod model to study the effects of the social influence between agents. Our approach is
similar, but not exactly equal, to that proposed in Ref. \cite{AFlache08082710}. In our case, at a given time $t$ an
agent $i$ is randomly chosen for possible influence. Around any agent, we consider a neighborhood defined by a
parameter $p$, which controls how many agents can be visited (in unitary steps) in each direction starting from that
agent until the perimeter of the neighborhood is reached. One unit-step allows to move only in north-south or east-west
direction. When $p$=1 the neighborhood obtained is therefore the well-known von Newmann neighborhood $N_1$, which only
includes the four closer neighbors of an agent ($N_1 = \{j_1, j_2, j_3, j_4$\}). In our case, $p$ can take higher
values to define bigger neighborhoods. In general, for a given value of $p$ it is obtained $N_p = \{j_1, ...,
j_\alpha\}$ where $\alpha = \alpha(p) = 2 p (p + 1)$. This give four neighbors ($\alpha = 4$) for one step neighborhood
($p=1$), twelve neighbors ($\alpha = 12$) for two step neighborhood ($p=2$) and so forth.

The social influence is defined as follows: when an agent $i$ is selected for possible influence, all of its neighbors
(according to the value of $p$) are included in the set of influence I$_i$ with probability $p_{ij_\alpha} =
O_{ij_\alpha} / F$, where $O_{ij_\alpha}$ is the cultural overlap between agent $i$ and $j_\alpha$. The set I$_i$
obtained by this procedure becomes a set of traits' influence over agent $i$'s traits. In our model we have decided to
include also agent $i$ into the set I$_i$ with probability $p_{ii}=1$. This allows a self-reflection of agent $i$ about
its own traits when comparing with the traits of its neighbors. Then, each agent is aware of its own traits when being
influenced by its neighbors and its own traits count when deciding to change, or not, its corresponding value for a
different one.

\begin{figure} 
\vspace*{1cm}
\psfig{file=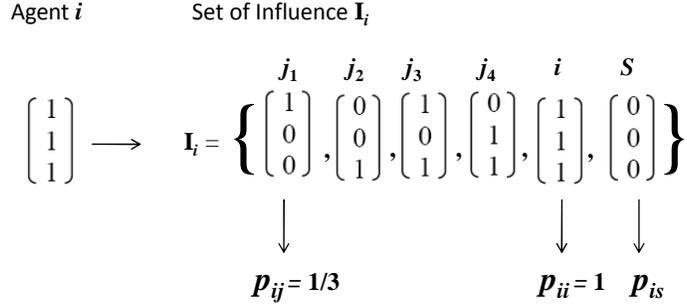,width=140mm}
\vspace*{-5cm}
\caption{Social interaction between agent $i$ and its neighbors. Each neighbor is included with probability
$p_{ij_\alpha}$ into the set of influence I$_i$. The agent $i$ is also included with probability $p_{ii}=1$. The agent
$s$, included with probability $p_{is}$ according to Eq. (\ref{ps}) represents a super-agent which acts over the whole
society.}
\label{fig0}
\end{figure}

Figure \ref{fig0} illustrates the social influence procedure. In the example shown, the agent $i$ selected has cultural
traits (1,1,1) and the neighborhood is formed by the one-step neighbors (von Neumann neighborhood). We have also
supposed that all the four neighbors have entered into the set of influence I$_i$. In general, neighbors are included
according to the probabilities $p_{ij}$, computed for each neighbor. The last agent to be included on the set I$_i$ is
an agent shared by all the agents on the network. Without loss of generality this super-agent $s$ is set up with
culture zero in all of its trait values and never change them. This global agent represents a Mass Media influence
which acts over the whole society. Its inclusion in this model follows the same rules explained on Ref.
\cite{arezky-ijmpc2009}, according to which the probability of interaction between this super-agent with an agent $i$
of the society is given by
\be
\label{ps}
p_{is} = \frac{O_{is} + \epsilon}{F + \epsilon}
\ee
where $O_{is}$ is the overlap between the agent $i$ and the super-agent $s$. The parameter $\epsilon (> 0)$ can take
real values and it is a measure of the Mass Media strength to act over all agents in the society. Then, the super-agent
$s$ has always a non-zero probability (min($p_{is}$) = $\epsilon/(F \, + \, \epsilon)$) to be included in the set of
influence of the agent $i$. This probability is proportional to the strength parameter $\epsilon$ and its non-zero
value means a clever Mass Media which designs its publicity actions to have always something in common with the target
of influence.

Once the agent $i$ is randomly selected and the current influence set I$_i$ is established (including $i$ and $s$), an
agent $i$'s feature is selected at random with a trait value different from the corresponding value of at least one
agent of the set I$_i$ in the same feature. The decision of which trait value agent $i$ adopts (if any change takes
place) is done in such a way that agent $i$ imitates (or copy) the most common trait in the set I$_i$ for the feature
selected. If, for example, in the situation depicted in Fig. \ref{fig0}, the feature selected on agent $i$ were the
second (that in the middle), this trait value ($\sigma_{i,2}=1$) will be changed to $\sigma_{i,2}=0$ due to highest
frequency of this last value in the set I$_i$. If the feature selected were the third (from top to down), the value
$\sigma_{i,3}=1$ will remain due to the highest frequency of this value on the set I$_i$. The last possible case
represented in Fig. \ref{fig0} occurs if the first feature were selected. In this case, on the set of influence I$_i$,
we have the same frequency of appearance for the values 0 and 1. In this situation, one of the two values is chosen at
random and then, the trait value $\sigma_{i,1}=1$ in agent $i$ could remain with probability $1/2$ or changed to
$\sigma_{i,1}=0$ with the same probability. Then, in general the trait value of agent $i$ is changed to the value of
the corresponding trait with the highest frequency of appearance on I$_i$. If there are more than one, it is changed to
one of them with equal probability. The trait value of agent $i$ does not change if this is the only one with the
highest frequency of appearance on I$_i$.

At the same time, we have also included {\it interaction} errors as well as {\it copying} errors, just as in Ref.
\cite{AFlache08082710}. The interaction error relaxes the previous deterministic procedure used in former works of
Axelrod model when deciding the possible interaction between two agents (dyadic interaction). Both copying and
interaction errors can randomly alter the outcome of an interaction event. The interaction error acts over the
selection procedure as follow: if the normal procedure of selecting an agent $j$ for being included in the set of
influence I$_i$ results in its inclusion, then with probability $r'$ the neighbor $j$ is removed from I$_i$. If the
neighbor has not been selected into the influence set, then with probability $r'$ the agent $j$ will be included into
I$_i$. This error creates the possibility that cultural influence occurs across the boundary of two disconnected
cultural regions if a neighbor with zero overlap is included in the influence set. This error can also reduce the
social pressure against adopting different trait values when a cultural identical neighbor is excluded from the set of
influence and increases the possibility for the target agent $i$ to adopt trait values from a completely different
culture. On the other hand, the copying error acts after agent $i$ has adopted (or not) a different trait value and the
new trait value has been already set up. In this case, the corresponding trait value adopted by agent $i$ is changed to
a new randomly selected value with probability $r$. Notice that the new value randomly generated could be that one the
agent $i$ already has. In general, both interaction and copying errors are conceptually different, but for simplicity we have used here the same value $r=r'$. Additionally we have implemented periodic boundary conditions (toroidal society) to avoid boundary effects.

In the case of dyadic interaction and in the absence of noise, when analyzing the possible final absorbing states, it
is simpler to establish it by checking when each agent has full or null overlap with each one of its neighbors. In the
case of social influence the problem is more involved. However, it is also possible to establish some technical
conditions to be checked to see if an agent is active (i.e., can interact with its neighbors according to the dynamical
rules established in the model) or not taking in consideration the set of influence I$_i$. When all the agents of the
society are inactive, then an absorbing state is obtained.

In the present study we have also implemented the procedure developed in Ref. \cite{fontanari-1} where a list of active
agents is built. In this way, instead of randomly selecting agents of the society in each time step, the agents are
randomly chosen directly from this list. This procedure strongly increase the efficiency of the dynamical evolution of
the system and it allows to save computational time. Therefore, when the system is initialized by randomly assigning
different cultures to each agent of the society, the first list of active agents is built. Next, at each time step,
when the influence is established and an agent of the society changes its cultural value, the list of active agents is
updated analyzing the agent itself and all its neighbors to check which of them are now active. The dynamical iteration
keeps on until the length of the list of active agents is reduced to zero. It is worth noticing that in our case (where
social influence is established), some runs did not settle into a final, well-defined absorbing state. In these cases
the list of active agents reduces to one element and each time this agent changes its cultural values and becomes
inactive, one of its neighbors becomes active. This propagation goes on indefinitely. We have neglected these cases
from our calculations and we have only considered runs which finish in a precise well-defined final absorbing state.
Moreover, when noise is included, it is not possible in general to define a final absorbing state and the dynamical
evolution of the system is stopped by defining some criterion related with a definition for the stationary state the
system reaches. In our case, we have included noise only in those agents which are active on each time step, and do not
in the rest of the society. This has allowed us to reach final absorbing states even when noise is present.

\section{\label{results} Results}

\begin{figure}
\vspace{-1cm}
\psfig{file=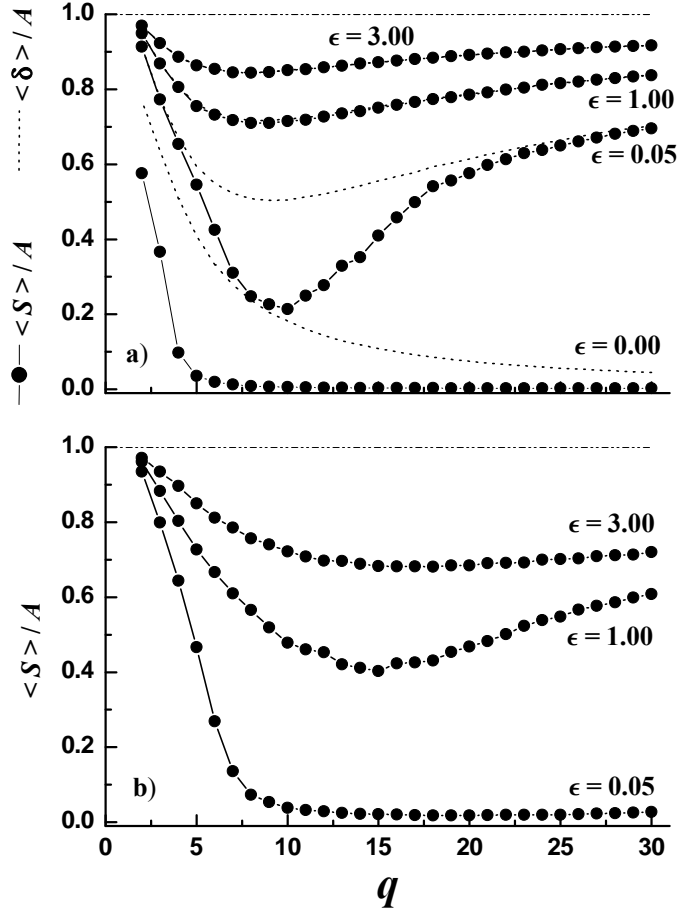,width=100mm}
\vspace{-1cm}
\caption{Final absorbing state in the Axelrod model with social influence as a function of $q$ for different values of
$\epsilon$. In dots-symbol line is the fraction of the biggest culture while in dotted line is the fraction of Mass
Media information on the society (only in panel a)). The parameter $A$ is the area of the system ($A=L^2$). a)
Calculation done including one-step neighbors (von Neumann neighborhood). b) Calculation done including two-step
neighbors. The calculation is done with $F$ = 3, $L$ = 50 and averages are taken over 200 different random initial
conditions.}
\label{fig1}
\end{figure}

We have implemented different computational experiments to study the Mass Media effects in the Axelrod model when the
social influence is the mechanism at work for agents interaction. As there is no qualitative difference on the
dynamical behavior of the system for $F \ge 2$, we have set $F$ = 3 in our study, a network size with $L$ = 50 (2500
agents unless otherwise stated) and averages have been taken over 200 different initial random configurations.

In Fig. \ref{fig1} we represent the normalized value of $S$ (the amount of agents in the biggest culture domain when
the system reaches the final absorbing state), as a function of the initial diversity $q$ with no noise ($r$ = 0).
Panel a) corresponds to calculations with one-step neighborhood ($p$ = 1) while in panel b) the calculation was done
for $p$ = 2. This means that the amount of neighbors of each agent in panel a) are only four plus the super-agent while
in case b) each agent has now twelve neighbors beside the super-agent. Different values of the super-agent strength
$\epsilon$ are included. It is also included in dotted-line the parameter $\delta$ which is the amount of information
of the super-agent on the society. This parameter is obtained counting along all of agents and their corresponding
traits if the information is equal to that of the super-agent in the same feature. The final count is divided by $F$ to
scale the value in a way that can be shown in the same scale as the parameter $S$.

It can be seen that for low values of $q$ the system reaches a close-monocultural state for any value of the strength
$\epsilon$. These are rather trivial cases because these extremely low values of $q$ mean a very low initial diversity
of the system and a final close-monoculture state is then expected.

A monoculture state is also induced for high enough values of $q$ but the system is now more sensitive to the values of
$\epsilon$ than in the region of low $q$ values. In this case, higher values of $\epsilon$ induce a stronger
multicultural final state given by higher values of $<S>$. To explain this result, we note that higher values of $q$
mean there are initially a higher degree of cultural diversity on the society and this is reflected in sets of
influence I with low number of neighbors, i.e., each set of influence I$_i$ will be frequently composed by the own
agent $i$ and by the super-agent $s$, and with low probability by the neighbors since they probably do not share any of
their trait values with agent $i$. The probability of the super-agent to be included in the set of influence I
increases for increasing value of $\epsilon$. For higher $q$ values, in cases where I$_i$ is formed by agent $i$ and
super-agent $s$, the last one will be able to introduce its own value on agent $i$ with probability 0.5. The iteration
of this process in time drives the system to a close monocultural state.

A more interesting situation occurs for middle values of $q$ where it is observed a minimum of the $<S>/A$ values as a
function of $q$, more pronounced for $\epsilon$ = 0.05 and $q$ = 10 in Fig. \ref{fig1} a) and $\epsilon$ = 1.00 and $q$
= 15 in Fig. \ref{fig1} b). For values of $q$ close to the corresponding minimum, the initial diversity is such that
besides agent $i$ and $s$, some of the agent's neighbors are also included in the set of influence I. Then the
interacting dynamics involves higher cultural diversity on the set of influence I$_i$ and the super-agent fails to
induce a strong monocultural final state.

It is interesting to note that when comparing Fig. \ref{fig1} a) and b) it can be seen that the increase of the number
of neighbors included in the social interaction decreases the size of the biggest culture in the absorbing state and
then a more pronounced multicultural state is reached for the same values of $q$ and $\epsilon$. The value of $q$ where
the minimum of $S$ is attained also increases. This is a consequence of a direct competition between the higher
diversity on the set I$_i$ and the homogenized influence of the super-agent. In Fig. \ref{fig1} b), the amount of
neighbors to be analyzed for inclusion on the set I$_i$ is twelve. Three times the case in Fig. \ref{fig1} a), which is
only four. Then with $p$ = 2, at any value of $q$ there will be, with higher probability, bigger diversity of trait
values than in the case of $p$ = 1 and, therefore, a weakening of the homogenizing effects of the super-agent is
expected. This decreases the possibility of the super-agent to drive the system to a monocultural absorbing state
because the probability for a trait value of the super-agent to appear with the highest frequency on the set of
influence is lower and lower values of $S$ are obtained. Then, when considering more neighbors in the social
interaction the higher local diversity reinforces the final multicultural state, even with the presence of a mass
media.

On the other hand, when comparing the dots line for $<S>$ with the dotter line for $<\delta>$ in Fig. \ref{fig1} a) it
can be seen that they both coincide for all values of $q$ when $\epsilon$ has higher values ($\epsilon = 1.00$ and
$\epsilon = 3.00$). It means that the information of the super-agent is present only in the biggest culture of the
society. Nevertheless, for small values of $\epsilon$, the two curves coincide for higher values of $q$, but the values
of $<\delta>$ are higher than the values of $<S>$ for middle values of $q$. Then, in this situation although the final
state does not correspond to a strong monoculture, the information of the super-agent on the society include the bigger
culture and also agents which belong to smaller culture sizes.

\begin{figure}
\vspace{-1cm}
\psfig{file=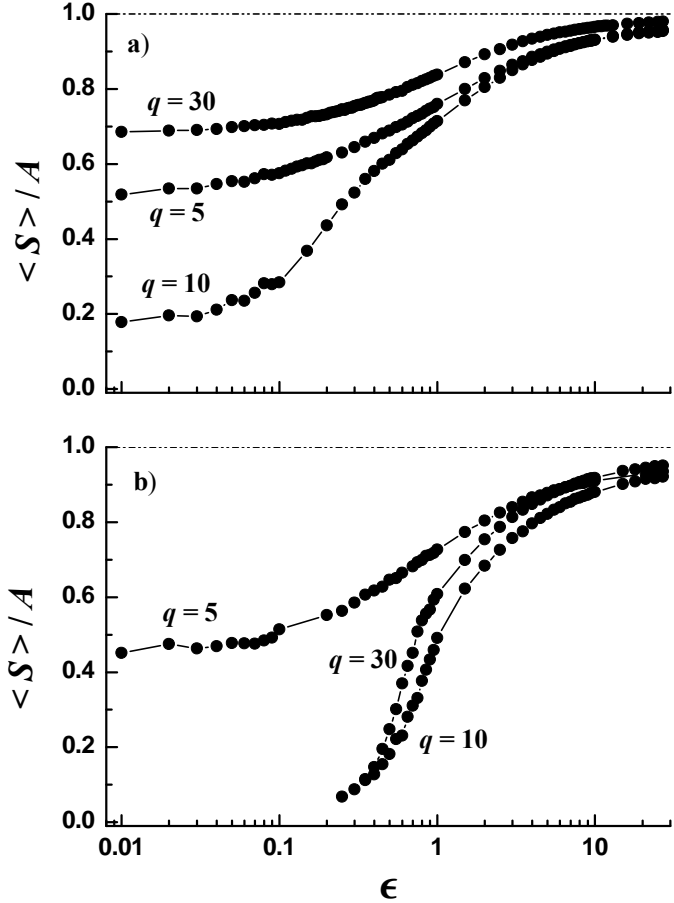,width=100mm}
\vspace{-1.5cm}
\caption{Final absorbing state in the Axelrod model with social influence as function of the Mass Media strength
$\epsilon$ for different values of $q$. a) Calculation done including one-step neighbors (von Neumann neighborhood). b)
Calculation done including two-step neighbors. The parameters used for these calculations are the same as in Fig.
\ref{fig1}.}
\label{fig2}
\end{figure}

Figure \ref{fig2} shows a second computational experiment. Here, we have included the values of the same parameter
$<S>/A$, but now as a function of the super-agent strength $\epsilon$ for three different values of $q$. In this case,
noise has also been neglected. One and two-steps neighborhoods are included in panels a) and b) respectively. For the
three values of $q$ shown, a monoculture is strongly induced as $\epsilon$ increases due to the constant presence of
the super-agent on the set of influence. The order of the curves according to the value of $q$ is in correspondence
with the order of the curves in Fig. \ref{fig1} and the values of $\epsilon$. As $\epsilon$ increases the super-agent
becomes a factor of ``normalization'' making constant its own values in time along the dynamical evolution of the
system. Even for $q$ = 10, when the set of influence includes with high probability some of the neighbors of the target
agent and the super-agent traits values are not on the majority frequency of appearance, the latter succeed in
inducing a monoculture as the strength $\epsilon$ increases. In particular, for $\epsilon \geq 1$, an almost
monocultural state with $<S>/A \geq 0.7$ is already induced, i.e., at least 70$\%$ of the society belongs to the
biggest culture for all values of $q$ included. For $\epsilon$ = 1, we have min($p_{is}$) = 0.25 (this is obtained
considering null overlap between the super-agent $s$ and an agent $i$) but the super-agent can induce the 70$\%$ of the
society to belongs to the bigger culture with its own information (according to the value of the parameter $\delta$
shown in Fig. \ref{fig1}), for any initial diversity $q$. The maximum value of $\epsilon$ included on Fig. \ref{fig2}
is $\epsilon$ = 27, which gives min($p_{is}$) = 0.90. Furthermore, when comparing in Fig. \ref{fig2} the panels a) and
b) the same results are obtained as when comparing in Fig. \ref{fig1} a) and b). In this case the inclusion of a
two-step neighbors in the social influence dynamics of the system strongly decreases the value of $S$ for the case $q$
= 10 and 30.

\begin{figure}
\vspace{-1cm}
\psfig{file=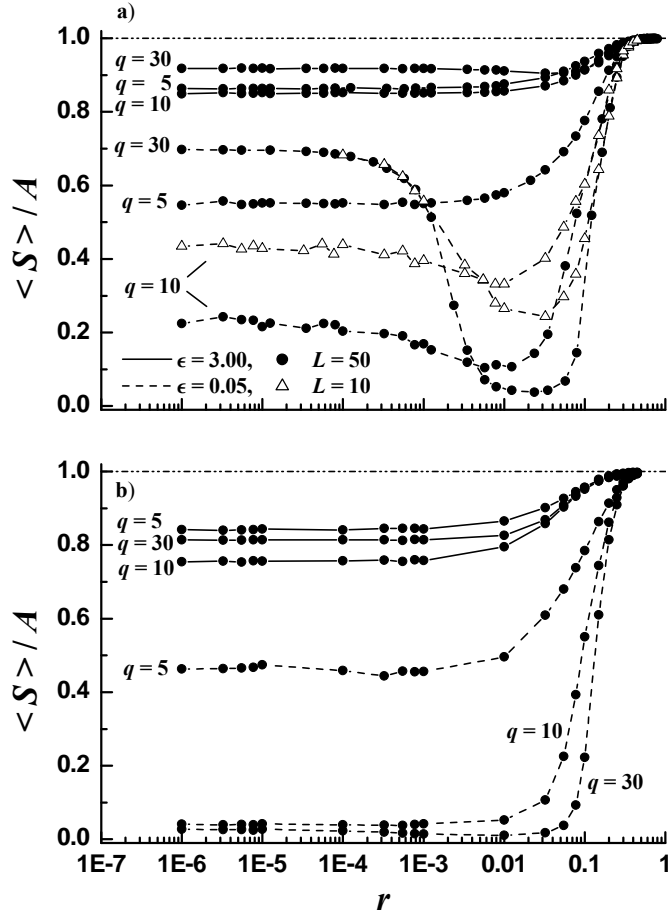,width=100mm}
\vspace{-1.5cm}
\caption{Dependence of the final absorbing state with respect to different noise levels and different values of $q$.
In solid line it is included the case with $\epsilon$ = 3.00 while dashed lines are the cases with $\epsilon$ = 0.05.
In black circles are reported the final absorbing states calculated with a network size of $L$ = 50, while in white
triangles the case with $L$ = 10. a) Calculation done including one-step neighbors (von Neumann neighborhood). b)
Calculation done including two-step neighbors.}
\label{fig3}
\end{figure}

In order to get additional insights on the system dynamics, we next consider the same model but including noise. In
Fig. \ref{fig3} it is reported the dependence of the parameter $<S>/A$ with respect to different noise levels for
several values of $q$, $\epsilon$ and $L$. In panels a) and b) we have considered one- and two-steps neighborhoods,
respectively. The noise included ranges between 0 and 0.45. As can be seen, social influence makes absorbing states
more stable to a bigger range of noise level than dyadic interaction, as already reported in Ref.
\cite{AFlache08082710}. In general there is no qualitative change for at least three order of magnitude for all the
values of $q$, $\epsilon$ and $L$ used. For higher strength of Mass Media ($\epsilon$ = 3.00, min($p_{is}$) = 0.5) the
final absorbing state remains almost monocultural for four orders of noise level and finally for noise values higher
than 0.1 the system is driven to a full monocultural state reached at $r$ = 0.45 approximately. The situation is
different for low values of super-agent strength ($\epsilon$ = 0.05). In this case the final absorbing state is also
stable to different noise levels, but only for three orders of magnitude. In panel a) of the Figure it can be seen that
for $q$ = 5 and noise level higher than $r$ = 0.01, the system is driven to a monoculture. For increasing noise level
at $q$ = 10 and 30 (for $\epsilon$ = 0.05) there is first a reinforcement of the multicultural state, given by
decreasing values of $<S>/A$ (stronger induced for $q$ = 30), and later for higher values of $r$ it is induced a
monoculture. All the calculations shown in black circles have been done with a network size of $L$ = 50. In order to
study if the stability of the final absorbing state to noise is robust with respect to the network size, we have also
explored the case $L$ = 10. Quantitative differences arise only for $q$ = 10 and 30, for $\epsilon$ = 0.05 (white
triangles in Fig. \ref{fig3} a)). In general it is observed that the final absorbing state is stable to the same range
of noise independent of the network size. The difference obtained, consisting in the higher values of $<S>/A$ for $q$ =
10 and the higher minimum for $q$ = 30 at $r$ = 0.05 (both for $\epsilon$ = 0.05) are due to the lower size of the
network ($L$ = 10). There are no quantitative differences for the other parameters due to the rather trivial case of
$q$ = 5 (for any value of $\epsilon >$ 0) and for the high value of $\epsilon$ used in the solid lines.

In Fig. \ref{fig3} b) we have included the same results of panel a) calculated with $L$ = 50, but now considering
two-step neighbors. The results are qualitatively the same for $\epsilon$ = 3.00 (no matter the values of $q$) and for
$q$ = 5 and $\epsilon$ = 0.05. For the case of $\epsilon$ = 0.05 and $q$ = 10 and 30 the system ends in a multicultural
state for almost all the range of noise levels. Nevertheless, for sufficiently high noise level the system is abruptly
driven to a monocultural state, as in all the others cases.

Here we advance an analysis for possible explanation of the results obtained on this experiment. There are three
parameters involved: the initial cultural diversity $q$, the strength of the super-agent $\epsilon$ and the noise level
$r$, and two different error mechanisms: the copying and including error. For low values of $\epsilon$ the super-agent
will appear on the set of influence I$_i$ with low frequency. This frequency increases for increasing $\epsilon$. For
low values of $q$ the majority of the neighbors of the target agents will be included on I$_i$ because of the low
cultural diversity, while the opposite occurs for large enough values of $q$. Nevertheless, as can be seen from
the Figure, this complex interplay does not has an important impact on the dynamics of the system along a wide range of
noise levels and the value of $<S>$ remains almost constant. When $r \geq$ 0.1 the copying error becomes the dominant
mechanism because independently of the cultural trait an agent has copied from its neighbors, the copying error changes
it to any one randomly selected and as it was already said, it deletes the boundaries between different cultural
regions. It is also of importance that the error mechanism stops when the agent becomes inactive. Hence, the social
influence is the mechanism which allows an agent to be active/inactive and it is also the mechanism which switch on/off
the copying error. Then, as the copying error connects two completely different cultures the homogenization is favored
and agents become inactive (and noise stops) when its culture is completely equal to its neighbors. Therefore the final
monoculture obtained at $r$ = 0.45.

At lower values of noise a non monotonic behavior is obtained for $q$ = 10 and 30 when the strength $\epsilon$
has low values ($\epsilon$ = 0.05 in this case). The value of $<S>$ first decreases for increasing noise reinforcing
the multicultural state. This effect is much pronounced at $q$ = 30. In these cases, the initial diversity makes
that the set I$_i$ be formed only by the target agent and some of its neighbors. More neighbors will be present on the
set of influence for $q$ = 10 than for $q$ = 30. For increasing noise the including error makes the set I$_i$ populated
by both agents from cultures which share some traits with the target agent and also with agent neighbors which do not.
The super-agent will also be included. This interplay seems to produce strong local convergence and it drives the
system to a multicultural state for a range of noise between 0.01 and 0.1. Finally, as already noted, the copying error drives the system
to a monoculture state for even higher values of $r$.

\begin{figure}
\vspace{3cm}
\psfig{file=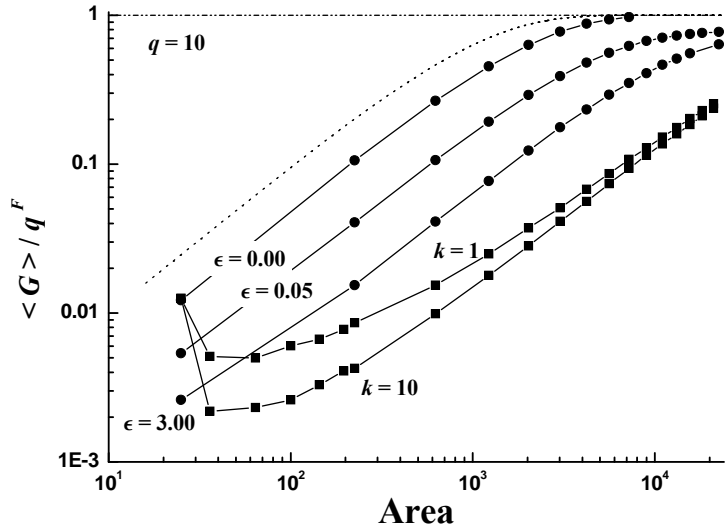,width=120mm}
\vspace{-3cm}
\caption{Culture-area relationship for $q$ = 10 and different values of $\epsilon$. In circles are included the
calculations for the corresponding value of $\epsilon$ indicated with labels. In square are represented the
calculations done according to the relation $\epsilon = k (L - 5)$, with the value of $k$ indicated with a label. $F$ =
3, $L$ = 50 and 200 different random initial conditions were used. The area is obtained according to $A = L^2$.}
\label{fig4}
\end{figure}

Finally, we study the influence of the Mass Media in the culture-area relation when the interaction on the society is
social. An extensive examination of the culture-area relation on the Axelrod model with dyadic interaction was done in
Ref. \cite{fontanari-1}. The authors obtained a non-monotonic behavior for the culture-area relation for $q$ values
below the critical value where the transition occurs ($q < q_c$) while for $q > q_c$ the number of cultures $G$ first
increases in a power-law dependence as $G \sim A^x$ with $x$ = 1 and then gradually flattens when the area becomes of
the order of the maximum number of cultures $q^F$. In the limit $L \rightarrow \infty$ there are only two possible
outcomes: for $q < q_c$ a single culture dominates in an ordered regime, while $G \rightarrow q^F$ and all the cultures
are represented in the network in a full disordered regime. The authors proved the transition between these two regimes
to be discontinuous because $G$ jumps from 1 to $q^F$ at $q = q_c$. Contrarily, when the model is changed including
instead of dyadic interaction social influence, Ref. \cite{AFlache08082710}, the problematic result of the Axelrod
model which produces a multicultural final state only for small societies \cite{fontanari-1} is overcome and only
multicultural final states are obtained. In case the Mass Media is included according to the model used in Refs.
\cite{avella2005,avella2006} and \cite{avella2007} which take the interaction between agents as dyadic, it was proven
in Ref. \cite{fontanari-2} that even a vanishingly small value of the Mass Media probability to interact with agents is
sufficient to destabilize the homogeneous regime for very large lattice sizes, contrary to the usual commonsense.

Here we have calculated the amount of different cultures present in the final absorbing state indicated by the
parameter $G$. We have followed the definition of Ref. \cite{fontanari-1} where all the different culture
configurations obtained are counted without paying attention to simple connected regions. As there are $q^F$ different
possible cultures, the parameter $G / q^F$ is normalized and its maximum value ($G / q^F$ = 1) means a completely
multicultural state while low values are related with monocultural final states.

In Fig. \ref{fig4} it is included with circles the final cultural diversity $G$ for a fixed value of the Mass Media
strength $\epsilon$. It can be seen that at $\epsilon$ = 0 the amount of cultures obtained in the final absorbing state
increases for increasing value of the area and reaches the maximum multicultural state for sufficiently high value of
$A$. The calculation done with any other value of $q$ gives the same dependence of the area which includes the same
slope and only a parallel shift. The other curves were then not included for simplicity. We have then qualitatively
reproduced the results obtained in Ref. \cite{AFlache08082710}. In our case, a calculation of the slope yields $x =
0.42 \pm 0.02$, which is different to that reported in Ref. \cite{fontanari-1}. We think that the deviation is due to
the difference that arises when dyadic interaction or social influence is included.

When the super-agent is included the values of $G$ are lower for higher strength $\epsilon$ at the same value of the
area, but the slope of the curve is the same as when the super-agent is not present. The super-agent also prevents the
system to reach the full multicultural state ($<G>/q^F = 1$) for higher values of the area. This can be seen on the
curve for $\epsilon$ = 0.05, which saturates at values of $A > 10^4$. The same seems to occur with $\epsilon$ = 3.00.
Calculations taking in consideration two-step neighbors were also carried out with the same qualitative results
(including slope and saturation). Only a shift to decreasing values of $G$ were obtained for the case of $\epsilon$ =
0.00 and 0.05, but the shift finally disappears for $\epsilon$ = 3.00 meaning that for strong enough Mass Media the
relative importance of the amount of neighbors including in the set of influence is weak (at least for the value of $q$
= 10).

Furthermore, on the dynamics of the system there are two parameters which compete to produce opposite effects for
increasing values of each one. For increasing area of the network the system tends to reach a multicultural state while
an increasing value of the super-agent strength $\epsilon$ pushes the system to a more cultural homogeneous absorbing
state. In order to study the relative weight these parameters have over the system, we have also calculated the final
absorbing state when both area $A$ and strength $\epsilon$ increases. To accomplish this purpose we have established a
dependence of the strength $\epsilon$ to the area given by the following relation:
\be
\label{ep-A}
\epsilon = k (A - 5)
\ee

In Fig. \ref{fig4} it is included, with square symbols, the results for $k$ = 1 and 10. The values of the diversity $G$
obtained are higher for $k$ = 1, than for $k$ = 10, due to the lower rate of increase of the strength $\epsilon$ and
consequently a more multicultural final state is induced. Nevertheless, both curves increase for increasing area, which
means that the area of the network has more weight than the strength $\epsilon$ on the dynamics of the system, as
expected if we examine Eq. (\ref{ps}). In our case, the maximum overlap between the super-agent and any agent of the
network is max($O_{is}) = F = 3$. Then, for $\epsilon/F \gg 1$ (which occurs rapidly for increasing value of $A$ in Eq.
\ref{ep-A}) it is obtained than $p_{is} \longrightarrow 1$ and the relative increase of the probability $p_{is}$ is
cancelled out with the effects of the increasing area. This explains also why the curves for $k$ = 1 and 10 tend to the
same values as the area is increased.

Then, when social interaction is present according to the present model, increasing network size always drives
the system to a multicultural state, while increasing super-agent strength prevents the system to reach the maximum
possible of cultural configurations. Additionally, the saturation value of $G$ seems independent of the value of the
strength of the mass media, as far as our calculations have shown.

We have also included in dotter line in Fig. \ref{fig4} the analytical expression reported in Ref. \cite{fontanari-1}
\be
\label{analyt}
<G> = q^F \left( 1 - \e^{\textstyle{- A / q^F}} \right)
\ee
which is the average number of cultures in the totally disordered configuration were $A = L^2$ agents are randomly
assigned with one of the $q^F$ available cultures. The expression is valid for $A$ and $q^F$ very large. As can be
seen, the prediction from Eq. (\ref{analyt}) overestimates the values for the cultural diversity as a function of the
area. The results are more different the higher the values of the super-agent strength $\epsilon$ are, since it is a
factor that decreases the cultural differences between neighboring agents.

\section{Conclusions.}

Summing up, we have numerically studied an Axelrod model in which two different mechanisms are at work. The first takes
into account that agents are prone to change their state according to a sort of social influence given by a (variable)
neighborhood made up of their nearest (both in space and in cultural similarity) neighbors. The dynamics of the agents
is also ruled by a Mass Media effect, represented in this case by an agent that keeps fixed its cultural traits all the
time. The dynamics of the system is such that at low and high values of initial social diversity $q$, a monocultural
state is attained, with a stronger dependency on the mass media strength ($\epsilon$) at large $q$ values. The reason
behind the formation of these monocultural states is the varying size of the set of neighbors with social influence:
the higher the diversity of initial cultural traits, the less is the number of agents in the neighborhood of an agent.
Admittedly, the more interesting behavior is for intermediate values of $q$, where the system dynamics attains a
minimum in the size of the biggest monocultural cluster. At those values of $q$ the initial diversity is such
that the mass media has to compete with a larger number of neighbors of the agent and thus its information is not necessary on
the majority. Then, in this case the mass media fails to drive the system to an homogeneous cultural state.
Nevertheless, the increase of the mass media strength for a fixed network size always reinforces the monocultural
state.

On the contrary, the increase of the network size was proven to be a parameter which strongly drives the system to a
polarized society where all the possible cultural configurations are present. The amount of final number of cultures
also follows a power-law dependence when the social influence is at work, but the exponent value found here is lower
than that reported for the case of the original Axelrod model with dyadic interaction. When mass media is present it
only succeeds in preventing the full multicultural state for increasing size of the society and the amount of cultures
obtained saturates. Moreover, the saturation value was found to be independent of the mass media strength in our
calculations.

We have also studied the model with noise representing errors in copying traits and in the formation of the social
influence interactions. In radical difference with previous works where the noise is present on the whole society, we
have included here noise effects only in those agents which are active according to the rules of the social
interaction. This allowed us to reach well-defined final absorbing states and correspondingly a higher precision in the
description of the noise effects. The results show that social influence makes the system dynamics more stable against
the presence of noise and that the latter only has a marginal influence on the general qualitative picture obtained
without any errors. It is however worth stressing that noise has in general a positive effect in the formation of
monocultural final states, giving rise to polarized societies for large enough values of it, independently of the other
parameters ruling the size of the social influence and the strength of the mass media effect. Nevertheless, the fact
that we have not distinguished between the two error mechanisms in this study does not allow to better understand the
specific role each one plays on the dynamics of the system and further studies are needed where these two types of
errors were considered as independents.

We also think about the need to conduct more detailed studies exploring the implications of the social influence with
respect to the dyadic one. The introduction of different models which better study the inter-relations between the
dyadic interaction and social influence on a social system could be of importance to further elucidate the robustness of
the results found here. This work is currently in progress.

\section*{Acknowledgements}

A. H. R. thanks to the International Center for Theoretical Physics (ICTP), were the major part of this work was
carried out. Y. M. is supported by Spanish MICINN through projects FIS2008-01240 and FIS2009-13364-C02-01.

\end{document}